%% file: proceedings.tex
\documentclass[a4paper]{jpconf}
\usepackage{graphicx}
\usepackage{amsmath}
\usepackage{hhline}
\usepackage{bbm}
\usepackage[table]{xcolor}

\input{texdefs}

\begin{document}

\title{Model building with the non-supersymmetric heterotic SO(16)xSO(16) string 
\footnote{LMU-ASC 07/15}}
\author{Stefan Groot Nibbelink}
\address{Arnold Sommerfeld Center for Theoretical Physics, 
Ludwig-Maximilians-Universit\"at M\"unchen, 80333 M\"unchen, Germany}
\ead{Groot.Nibbelink@physik.uni-muenchen.de}

\begin{abstract}
In this talk we review recent investigations~\cite{Blaszczyk:2014qoa} of  the non-supersymmetric heterotic SO(16)$\times$SO(16) string on orbifolds and smooth Calabi-Yaus. 
Using such supersymmetry preserving backgrounds allows one to re-employ commonly known model building techniques. 
We will argue that tachyons do not appear on smooth Calabi-Yaus to leading order in $\alpha'$ and $g_s$.
Twisted tachyons may arise on singular orbifolds, where some of these approximations break down. However, they get lifted in full blow-up. 
Finally, we show that model searches is viable by identifying over 12,000 of SM-like models on various orbifold geometries. 
\end{abstract}

\section{Introduction}
\label{sc:Introduction}

For the last two decades or so the conventional way to search for particle physics models from the heterotic string aimed to construct supersymmetric Standard Model(MSSM)-like models from the heterotic string theory. On smooth Calabi--Yau spaces with non--Abelian vector bundles the authors of~\cite{Candelas:1985en} have obtained MSSM--like models~\cite{Bouchard:2005ag} with possible supersymmetry breaking built in~\cite{Braun:2005ux,Braun:2005bw,Braun:2005nv}. A more systematic search for MSSM--like modes has been performed using line bundles~\cite{Anderson:2011ns,Anderson:2012yf,Anderson:2013xka}. 
Orbifolds~\cite{Dixon:1985jw,Dixon:1986jc,Ibanez:1986tp,Ibanez:1987pj} may also be used to construct MSSM--like models in the heterotic string context, see e.g.~\cite{string_compactification_phys_rept,Choi2006}. 
In Refs.~\cite{Buchmuller:2005jr,Buchmuller:2006ik,Lebedev:2006kn,Lebedev:2008un} MSSM--like models have been assembled on the toroidal ${\mathbbm Z}_{\text{6-II}}$ orbifold. 
Similar investigations have been performed on a variety of other orbifolds~\cite{Z2xZ4,Kim:2006hv,Kim:2007mt,Blaszczyk:2009in,Nibbelink:2013lua}; for a comprehensive overview of orbifold model building see~\cite{Nilles:2014owa}. 

In this talk we entertain the question, whether it is possible to construct non-supersymmetric particle physics models from string theory. This talk is based on the recent publication~\cite{Blaszczyk:2014qoa}. The main motivation underlying this work is that so far no hints for the existence of supersymmetry have been found in particle physics experiments, even though the LHC has been extensively looking for it. This raises the question: why do (did) people in the string phenomenological community belief strongly in supersymmetry.  Here are some of the standard arguments: 
\items{
\item hierarchy problem 
\item unification of gauge couplings 
\item dark matter candidate 
\item compelling extension of the Poincar\'e group 
\item gain computational control 
}

Particle physics experiments, like the LHC, only provides us with bounds on the scale of supersymmetry breaking. The higher these bounds are pushed, the more seriously one should consider the possibility, that supersymmetry might not be realized up to the Planck scale at all. However, if there is no supersymmetry up to the Planck scale, does it still make sense to consider string theory as a framework to study particle physics? Indeed, the standard lore wants that supersymmetry is predicted by string theory. However, this is simply not true: String theory only requires worldsheet supersymmetry, but not necessarily in target space ~\cite{Dixon:1986iz,Dixon:1986jc,AlvarezGaume:1986jb}.

There have been various works on non-supersymmetric models in string theory over the years. Dienes~\cite{Dienes:1994np,Dienes:2006ut} performs some statistical scans of non-supersymmetric free-fermion models~\cite{Shiu:1998he} to give some idea of the scattering of the value of the cosmological constant. The connection between non-supersymmetric free fermionic models~\cite{Kawai:1986vd}, the Ho\v{r}ava-Witten model and other dualities have been studied in~\cite{Blum:1997cs,Faraggi:2007tj}.  
A large set of non-supersymmetric models in four dimensions were constructed using a covariant lattice approach~\cite{Lerche:1986ae,Lerche:1986cx}.
Non-supersymmetric tachyon-free type-I/II orientifold models 
\cite{Sagnotti:1995ga,Sagnotti:1996qj,Angelantonj:1998gj,Sugimoto:1999tx,
Blumenhagen:1999ns, Aldazabal:1999tw,Moriyama:2001ge}
have also been constructed as rational conformal field theories~\cite{GatoRivera:2007yi,GatoRivera:2008zn}. 

Such non-supersymmetric models from string theory face some potentially very serious problems: 
\items{ 
\item
the spectrum might be tachyonic
\item 
Higgs mass will be quadratically divergent
\item 
the cosmological constant problem with an associated destabilizing dilaton tadpole
\item 
far less  practical computational control
}
As stated above many of these problems were precisely the motivation for supersymmetry in the first place. 
In this talk we would like to focus on the lost of computational control in particular.  This problem indeed appears to be overwhelming: Generic six dimensional compactifications have not been classified;  even the number of toroidal orbifolds is almost 29 million. Hence, it is unclear how one should select promising backgrounds for phenomenological studies. However, even more problematic, most practical techniques we to compute spectra and couplings rely heavily on supersymmetry: Consequently, in the generic non-supersymmetric context we hardly have tools to investigate whether certain backgrounds have any phenomenological potential.

When studying non-supersymmetric models, it seems odd, or at least unnecessary, to focus on supersymmetric backgrounds. However, as we will see in this talk, to have computational control, it turns out to be extremely useful to consider the non--supersymmetric heterotic SO(16)$\times$ SO(16) string on backgrounds, that would preserve supersymmetry themselves~\cite{Blaszczyk:2014qoa}. It helps to restrict the number of geometries to be considered to a more moderate and managable number. More importantly, on supersymmetric backgrounds many of the known computational techniques can be recycled even if the theory is non-supersymmetric itself. 

\subsection*{Overview}

In the remainder of this talk we first briefly introduce the non-supersymmetric heterotic SO(16)$\times$SO(16) string. After this introduction we describe compactifications of it on smooth Calabi-Yau spaces. We explain how spectrum computations can be amended to determine the complete massless spectrum for both chiral fermions and complex bosons. In addition, we argue that tachyons can always be avoided on smooth Calabi-Yaus to leading order in the various string expansions. After that we discuss orbifolds of the non-supersymmetric hetorotic string. We conclude the talk with a summary of our model building results on orbifolds.

\section{Non-supersymmetric SO(16)$\times$SO(16) heterotic string}
\label{sc:10DN=0}

\begin{table} 
\begin{center} 
  \renewcommand{\arraystretch}{1.2}
\begin{tabular}{|c | l |}
\hline 
Fields & Space-time interpretation 
\\ \hline\hline 
 \cellcolor{lightgray} $G_{MN}, B_{MN}, \phi$ 
 &  \cellcolor{lightgray}Graviton, B-field, Dilaton
\\
 \cellcolor{lightgray}$A_M$ 
 &  \cellcolor{lightgray} SO(16)$\times$SO(16) Gauge fields 
\\ \hline 
$\Psi_+$ & Spinors in the $(\mathbf{128},\mathbf{1})+(\mathbf{1},
\mathbf{128})$ 
\\
$\Psi_-$ & Cospinors in the $(\mathbf{16},\mathbf{16})$
\\ \hline 
\end{tabular}
  \renewcommand{\arraystretch}{1}
\end{center}
\caption{\label{tb:N=0Spectrum}
This table presents the massless spectrum of the ten-dimensional non-supersymmetric heterotic SO(16)$\times$SO(16) theory. We indicate the bosons and fermions with gray and white backgrounds, respectively, in this and subsequent tables.}
\end{table}

In order to better understand the four-dimensional non-supersymmetric models emerging from string theory, we take as our starting point the ten-dimensional non-supersymmetric heterotic SO(16)$\times$SO(16) theory~\cite{Dixon:1986iz,Dixon:1986jc,AlvarezGaume:1986jb}. Its low energy spectrum is given in Table~\ref{tb:N=0Spectrum}. It contains the graviton, B-field and dilaton but none of their supersymmetric partners. There are gauge fields associated to the gauge group SO(16)$\times$SO(16). But instead of their gauginos, the low energy spectrum contains matter spinor and cospinor states (fermionic states with opposite ten dimensional chiralities) in the representations $(\mathbf{1},\mathbf{128})+(\mathbf{128},\mathbf{1})$ and $(\mathbf{16},\mathbf{16})$ of the gauge group, respectively.

\begin{figure}
\begin{center} 
\scalebox{.6}{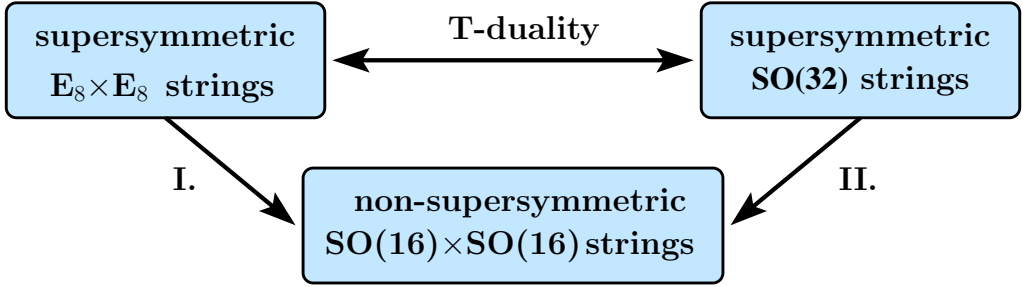}
\end{center}
\caption{\label{fg:RelationHetStrings}
This figure depicts the relation between the three heterotic string theories in ten dimensions.} 
\end{figure}

It is well-known that the supersymmetric E$_8\times$E$_8$ and SO(32) heterotic strings are dual to each other once compactified on a circle. The non-supersymmetric heterotic string is related to both these supersymmetric theories as well by orbifolding. The relations between the three consistent ten dimensional heterotic theories are indicated in Figure~\ref{fg:RelationHetStrings}: 

In more detail, the non-supersymmetric heterotic SO(16)$\times$SO(16) theory can be obtained by a (freely acting) $\Intr_2$ orbifolding of the supersymmetric E$_8\times$E$_8$ theory~\cite{Dixon:1986iz,Dixon:1986jc}. To this end the twist $v_0$ and the shift $V_0$ may be taken to be given by: 
\begin{equation}\label{NonSUSYE8}
v_0 = \big(0, 1^3\big)~,
\qquad 
V_0 = \big(1,0^7\big)\big(\sm1,0^7\big)~. 
\end{equation}
Even though this corresponds to trivial $2 \pi$ rotations on the three $\Real^2$ planes simultaneously, it act non-trivially on the target space fermions~\cite{Rohm:1983aq}.
Similarly, it can be obtained by considering a (freely acting) $\Intr_2$ orbifold of the supersymmetric SO(32) theory. This time the twist $v_0$ and the shift $V'_0$ may be  chosen as: 
\begin{equation}\label{NonSUSYSO}
v_0 = \big(0, 1^3\big)~,
\qquad 
V'_0 = \big(1,0^7\big)\big(\sm\sfrac 12, \sfrac 12^7\big)~. 
\end{equation}
These two different methods do not only give  the non-supersymmetric SO(16)$\times$SO(16) theory at the massless level, but, in fact, holds at the level of the full partition function and hence are relations between full string theories. An interesting observation is that the notion of which states in the SO(16)$\times$SO(16) theory one should refer to as twisted sector states, depends on which of the two supersymmetry heterotic strings one has started from: 

If one begins with the E$_8\times$E$_8$ theory, the states $(\mathbf{1},\mathbf{128})+(\mathbf{128},\mathbf{1})$ should be considered to be untwisted states as they are part of the ten-dimensional gauge multiplet, while the $(\mathbf{16},\mathbf{16})$ are twisted states. When one starts with the SO(32) string theory instead, their roles become precisely reversed. In particular, the $(\mathbf{16},\mathbf{16})$ states are untwisted in this case, as they arise from the branching of the adjoint of SO(32) to SO(16)$\times$SO(16). And, like in lower dimensional orbifolds of the SO(32) theory, the spinor representations only appear in twisted sectors.

\section{Smooth Calabi-Yau backgrounds}
\label{sc:Smooth}

To be able to recycle many of the computational techniques we have learned over the years in string phenomenology, we consider the compactification of the non-supersymmetric SO(16)$\times$SO(16) theory on backgrounds that would preserve supersymmetry themselves. This means, that we investigate the compactification of this theory on smooth Calabi-Yau manifolds with holomorphic vector bundles, subject to the integrated Bianchi identities 
\begin{equation} 
\int_{\mathcal{C}^4} \left\{ \tr\, \mathcal{R}_2^2 - \tr\, \mathcal{F}_2^2 \right\} = 0~, 
\end{equation} 
where  $\mathcal{R}_2$ is the curvature two-form and  $\mathcal{F}_2$ the gauge field strength, on any closed four-cycle $\mathcal{C}^4 \subset \mathcal{M}^6$  (see e.g.\ \cite{Candelas:1985en,gsw_2}). Let us list some advantages of this approach:

\subsection*{Fermionic spectrum}

In particular, to compute the spectrum of chiral fermions in four dimensions we can immediately recycle the use of cohomology and (representation-dependent) index theorems, like the multiplicity operator~\cite{Nibbelink:2007rd,Nibbelink:2007pn}
\begin{equation}  
\mathcal{N} = \int \left\{
\frac 16\, \Big(\frac{\mathcal{F}_2}{2\pi}\Big)^3 - \frac 1{24}\, \frac{\mathcal{F}_2}{2\pi}\, \tr\, \Big(\frac{\mathcal{R}_2}{2\pi}\Big)^2 \right\}~, 
\end{equation}
which is evaluated on all the fermions listed in Table~\ref{tb:N=0Spectrum} (keeping track of the ten-dimensional chirality).

\subsection*{Bosonic spectrum}

On generic six-dimensional manifolds it is difficult  to determine  the number of zero modes of the Laplace operator $\gD$. However, on a smooth Calabi-Yau manifold $\mathcal{M}^6$ with a vector bundle we can use that the Laplace operator for complex scalars,
\equ{ \label{Laplace} 
\gD \sim (iD\Slashed)^2~,
}
 is related to the Dirac operator $i D\Slashed$ of the would be gauginos. Hence, this allows us to use (representation dependent) indices and cohomology theory to determine the spectra of complex scalars.

\subsection*{No tachyons on Calabi-Yau backgrounds}

This method to compute the bosonic spectrum has an important consequence: 
To leading order there are no tachyons on smooth Calabi-Yau manifolds in the large volume approximation. On smooth Calabi-Yau backgrounds the Laplace operator $\gD$ is related to the square of the Dirac operator $i D\Slashed$. As its spectrum is non-negative, hence so is the spectrum of the Laplace operator $\gD$. The fact that the would be fermionic partners do not exist in non-supersymmetric theory is irrelevant for this conclusion, since the relation \eqref{Laplace} between the Laplace and the Dirac operator is a purely algebraic relation between these operators. 
Hence a major problem of non-supersymmetric model building, tachyons, can be avoided by working on smooth Calabi-Yaus to lowest order in $\alpha'$ and $g_s$ corrections.

\subsection*{Standard embedding gives on SO(10) GUT}

\begin{table}
\begin{center}
\renewcommand{\arraystretch}{1.2}
\scalebox{1}{
 \begin{tabular}{|c||c|c|}
 \hline
 Multiplicity & \cellcolor{lightgray} Complex bosons & Chiral fermions \\
 \hline \hline 
 1 & \cellcolor{lightgray} $-$ & $ (\mathbf{16};\mathbf{1})_{3}  + (\overline{\mathbf{16}};\mathbf{1})_{\sm3} + (\mathbf{1};\mathbf{128})_0 + (\mathbf{10};\mathbf{16})_0$ \\
 \hline
 $h^{1,1}$ & \cellcolor{lightgray} $(\mathbf{10};\mathbf{1})_2 + (\mathbf{1};\mathbf{1})_{\sm 4} $ & $(\mathbf{16};\mathbf{1})_{\sm 1} + (\mathbf{1};\mathbf{16})_{\sm 2}$ \\
 \hline
 $h^{1,2}$ & \cellcolor{lightgray} $(\mathbf{10};\mathbf{1})_{\sm 2} + (\mathbf{1};\mathbf{1})_{4} $ & $(\overline{\mathbf{16}};\mathbf{1})_{1} + (\mathbf{1};\mathbf{16})_{2}$ \\
 \hline
 $h^{1}(\text{End}(V))$ & \cellcolor{lightgray} $(\mathbf{1};\mathbf{1})_0 $ & $ - $ \\
 \hline
 \end{tabular}}
\renewcommand{\arraystretch}{1}
\end{center}
\caption{
In the standard embedding the SO(16)$\times$SO(16) theory compactifed on a smooth Calabi-Yau leads to an SO(10) GUT theory. 
\label{tab:genCYspectra}}
\end{table}

Finally, the simplest choice of a gauge background is the so-called standard embedding, in which the gauge connection is set equal to the spin connection. In the non-supersymmetric SO(16)$\times$SO(16) theory the resulting unbroken gauge group reads: SO(10)$\times$U(1)$\times$SO(16)$'$. Hence, we see that already the standard embedding leads to a promising GUT theory. The bosonic and fermionic spectra was computed in~\cite{Font:2002pq}, see Table \ref{tab:genCYspectra}.

\section{Orbifold compactifications}
\label{sc:Orbifold}

To describe orbifolds of the non-supersymmetric SO(16)$\times$SO(16) theory we may use its construction as a non-supersymmetric orbifold of the supersymmetric E$_8\times$E$_8$ theory. A $\Intr_N$ Calabi-Yau orbifold of the E$_8\times$E$_8$ theory is defined by the worldsheet boundary conditions: 
\equ{ 
X^i(\sigma+1) = e^{2\pi i k v_i}\, X^i(\sigma)~, 
\qquad
\psi^i(\sigma+1) = e^{2\pi i (\frac s2 + k v_i)} \, \psi^i(\sigma)~, 
\\[1ex] 
\lambda^I_1(\sigma+1) = e^{2\pi i (\frac {t}2 +k V_{1I})}\, \lambda^I_1(\sigma)~, 
\qquad 
\lambda^I_2(\sigma+1) = e^{2\pi i (\frac {u}2 +k V_{2I})}\, \lambda^I_2(\sigma)~. 
}
The $\Intr_N$ action is encoded in a twist $v$ and a gauge shift $V=(V_1;V_2)$ satisfying: 
\equ{
N\, v_i \equiv 0~, 
\qquad 
N\, V_{1,2} \in \text{weight lattice of }\mathbf{E}_8~. 
}
We focus on $\Intr_N$ orbifold twists that would preserve at least four dimensional, N=1 supersymmetry if applied to the E$_8\times$E$_8$ theory:
\begin{equation} 
v = (v_1,v_2, -v_1-v_2)~, 
\end{equation} 
like $(\sfrac 13, \sfrac 13, - \sfrac 23)$ for the $T^6/\Intr_3$ orbifold. 
We require that we have modular invariant partition function for the full orbifolded non-supersymmetric SO(16)$\times$SO(16) theory, e.g. 
\begin{equation} 
\frac {N}2 \,(V^2-v^2) \equiv V_0 \cdot V  - v_0\cdot v \equiv 0~, 
\end{equation} 
with the supersymmetry breaking twist and shift, $v_0$ and $V_0$, given in 
\eqref{NonSUSYE8}.

\subsection*{Twisted tachyons}

\begin{table}[]
\begin{center}
\renewcommand{\arraystretch}{1.4}
\scalebox{1}{
\begin{tabular}{|l|c|c||l|c|c|}
\hline 
\multicolumn{1}{|c|}{Orbifold} & Twist & Tachyons & \multicolumn{1}{|c|}{Orbifold} & Twist & Tachyons 
\\ \hline\hline 
$T^6/\Intr_3$ & $\frac13(1,1,-2)$ & forbidden &
$T^6/\Intr_2\times\Intr_2$ & $\frac12(1,-1,0)\,;~ \frac12(0,1,-1)$ & forbidden
\\ \hline 
$T^6/\Intr_4$ & $\frac14(1,1,-2)$ & forbidden &
$T^6/\Intr_2\times\Intr_4$ & $\frac12(1,-1,0)\,;~ \frac14(0,1,-1)$ & possible
\\ \hline 
$T^6/\Intr_\text{6-I}$ & $\frac16(1,1,-2)$ & possible &
$T^6/\Intr_2\times\Intr_\text{6-I}$ & $\frac12(1,-1,0)\,;~ \frac16(1,1,-2)$ & possible
\\ \hline 
$T^6/\Intr_\text{6-II}$ & $\frac16(1,2,-3)$ & possible & 
$T^6/\Intr_2\times\Intr_\text{6-II}$ & $\frac12(1,-1,0)\,;~ \frac16(0,1,-1)$ &  possible
\\ \hline 
$T^6/\Intr_7$ & $\frac17(1,2,-3)$ & possible &
$T^6/\Intr_3\times\Intr_3$ & $\frac13(1,-1,0)\,;~ \frac13(0,1,-1)$ & possible 
\\ \hline 
$T^6/\Intr_\text{8-I}$ & $\frac18(1,2,-3)$ & possible &
$T^6/\Intr_3\times\Intr_6$ & $\frac13(1,-1,0)\,;~ \frac16(0,1,-1)$ & possible
\\ \hline 
$T^6/\Intr_\text{8-II}$ & $\frac18(1,3,-4)$ & possible & 
$T^6/\Intr_4\times\Intr_4$ & $\frac14(1,-1,0)\,;~ \frac14(0,1,-1)$ & possible
\\ \hline 
$T^6/\Intr_\text{12-I}$ & $\frac1{12}(1,4,-5)$ & possible & 
$T^6/\Intr_6\times\Intr_6$ & $\frac16(1,-1,0)\,;~ \frac16(0,1,-1)$ & cpossible 
\\ \hline 
$T^6/\Intr_\text{12-II}$ & $\frac1{12}(1,5,-6)$ &  possible  
\\ \hhline{---}
\end{tabular}}
\renewcommand{\arraystretch}{1}
\end{center}
\caption{\label{tb:TachyonicOrbifolds}
This table indicates on which orbifolds of the non-supersymmetric heterotic SO(16)$\times$SO(16)  tachyons are not strictly forbidden. When tachyons are possible on a certain orbifold, this does not mean that all models on this space necessarily will contain tachyons.}
\end{table}

Since orbifolds can be quantized exactly, one can directly investigate whether tachyonic states may arise when compactifying the non-supersymmetric heterotic SO(16)$\times$SO(16) string on a Calabi-Yau orbifold. In Table~\ref{tb:TachyonicOrbifolds} some Abelian supersymmetric orbifolds have been listed. It is indicated which of those may have twisted tachyons. Moreover, the orbifolds, $T^6/\Intr_\text{12-II}$, $T^6/\Intr_2\times\Intr_\text{6-II}$, $T^6/\Intr_3\times\Intr_6$ and $T^6/\Intr_6\times\Intr_6$, even have tachyonic states with right-moving oscillators switched on. This table should be interpreted with care: When tachyons are possible, this does not mean that all such orbifold models actually contain tachyons; they may have been projected out.

\subsection*{Decoupling twisted tachyons in blow-up}

It sounds like, we have a conflict on our hands here: We have argued that the non-supersymmetric SO(16)$\times$SO(16) theory on smooth Calabi-Yaus to leading order do not have tachyons, while on orbifolds tachyons are possible even when they preserve supersymmetry. The point here is that an orbifold is a singular Calabi-Yau space, hence $\alpha'$ corrections cannot be ignored and hence our argument against tachyons breaks down. 

To investigate this in a bit more detail, we consider a $\Intr_\text{6-I}$ orbifold of the non-supersymmetric heterotic theory, with shift vector, 
\begin{equation} \label{TachyonicShiftExample}
V = \sfrac 16
\big( -2, -16,  -14,  -2,    2,     6,   3,  11\big) \big( -2,  -5,    -6,  -2,      6, -13,  -1,  19\big)~, 
\end{equation}
without any Wilson lines for simplicity. In the upper part of the Table~\ref{tab:TachyonicOrbifoldBlowup} we give the massless spectrum of this orbifold theory. In the lower part of this table we give the spectrum of a full resolution model using line bundles only. This spectrum we have determined using the multiplicity operator. Notice that all massless fields in blow-up can be matched with states in the orbifold spectrum taking into account possible field redefinitions and decoupling of vector-like states~\cite{GrootNibbelink:2007ew}. In particular, we observe that the tachyonic twisted states in the upper table have all disappeared in the lower one. 
\newpage

To understand how the twisted tachyons decoupled in the blow-up process, we zoom in on the following bosonic fields:  
\begin{center}
\renewcommand{\arraystretch}{1.2}
 \begin{tabular}{|cl|c|c|}
  \hline
\multicolumn{2}{|c|}{ State } & Sector & Representation \\
  \hline\hline  
  Tachyon & $t$ & $\theta^1$ & $(\rep1;\rep1,\rep1,\rep2)$ \\
  \hline 
  Blow-up mode & $b$ & $\theta^2$ & $(\rep1;\rep1,\rep2_-,\rep1)$ \\
  \hline 
  Complex scalar &  $c$ & $\theta^3$ & $(\rep1;\rep1,\rep2_-,\rep2)$ \\
  \hline
 \end{tabular}
 \renewcommand{\arraystretch}{1}
\end{center}
Following standard effective field theory intuition, which states that all interactions, which are not strictly forbidden, will arise, one anticipates  that the effective potential will be generated of the form: 
 \begin{equation} \label{EffPot} 
 V_\text{eff} = - m_t^2\, |t|^2 + |\lambda|^2\, |b|^2 \, |t|^2 + \ldots\,, 
 \end{equation}
where $m_t^2$ parameterizes the tachyonic mass. 
If the blow-up mode VEV is sufficiently big, the tachyon acquires a positive mass and hence decouples from the theory.

\begin{table}[t]
\begin{center}
\renewcommand{\arraystretch}{1.2}
\scalebox{.95}{
 \begin{tabular}{|c||c|}
 \hline
 States &  Gauge representations of the spectrum of a tachyonic $\Intr_\text{6-I}$ orbifold \\
 \hline \hline 
\cellcolor{lightgray} Bosonic tachyons & \cellcolor{lightgray}
  $3 (  \mathbf{1};   \mathbf{1},   \mathbf{1},   \mathbf{2})$
 \\ \hline\hline  
 Massless & 
 $ 4 (  \mathbf{10};   \mathbf{1}) +  (  \overline{\mathbf{10}};   \mathbf{1}) 
 + 6 ( \mathbf{5};   \mathbf{1})  + 3 ( \overline{\mathbf{5}};   \mathbf{1}) 
 + (   \mathbf{5};   \mathbf{1},   \mathbf{4},   \mathbf{1})
 + 2 ( \overline{ \mathbf{5}};   \mathbf{1},   \mathbf{1},   \mathbf{2})
 + ( \mathbf{5};   \mathbf{1},   \mathbf{1},   \mathbf{2})
 $
 \\
 chiral fermions & 
 $+ 2( \overline{ \mathbf{5}};   \mathbf{4},   \mathbf{1},   \mathbf{1})
 + 12 (\mathbf{1};   \mathbf{4},   \mathbf{1},   \mathbf{1})
 + 18  (\mathbf{1};   \overline{\mathbf{4}},   \mathbf{1},   \mathbf{1})
 +  2 (   \mathbf{1}; \overline{ \mathbf{4}},   \mathbf{2}_-,   \mathbf{2}) 
 + 2 (   \mathbf{1};   \mathbf{4},   \mathbf{2}_+,   \mathbf{1})
 $ 
\\ &
$+ ( \mathbf{1};   \mathbf{6},   \mathbf{2}_-,   \mathbf{1})
+ ( \mathbf{1};   \mathbf{6},   \mathbf{2}_+,   \mathbf{1})
+ 12 (\mathbf{1};   \mathbf{1},   \mathbf{2}_+,   \mathbf{2})
+ 4 (\mathbf{1};   \mathbf{1},   \mathbf{4},   \mathbf{1})
+ 36 ( \mathbf{1};   \mathbf{1},   \mathbf{2}_-,   \mathbf{1})$
\\ & 
$ + 30 (\mathbf{1};   \mathbf{1},   \mathbf{2}_+,   \mathbf{1})
+ 11 (\mathbf{1};   \mathbf{1},   \mathbf{1},   \mathbf{2})
+ 53 (   \mathbf{1};  \mathbf{1}) 
$
 \\ \hline\hline  
 \cellcolor{lightgray} Massless & \cellcolor{lightgray} 
$
9 ( \mathbf{5};   \mathbf{1})
+ 2 ( \overline{ \mathbf{5}};   \mathbf{1})
+ (\overline{ \mathbf{10}};   \mathbf{1})
+ ( \mathbf{1};   \mathbf{1},   \mathbf{4},   \mathbf{2})
+ 30 (   \mathbf{1};   \mathbf{1},   \mathbf{2}_-,   \mathbf{1})
+ 12 (  \mathbf{1};   \mathbf{6},   \mathbf{1},   \mathbf{1})$ 
 \\ 
\cellcolor{lightgray} complex scalars & \cellcolor{lightgray}
$+  2 (   \mathbf{1};   \mathbf{4},   \mathbf{1},   \mathbf{2})
+ 2 (   \mathbf{1}, \overline{ \mathbf{4}},   \mathbf{4},   \mathbf{1})
+ 22 (   \mathbf{1};   \mathbf{1},  \mathbf{2}_+,   \mathbf{1}) 
+ 10 (  \mathbf{1};  \mathbf{1},   \mathbf{2}_-,  \mathbf{2}) 
+ 46 ( \mathbf{1};  \mathbf{1})
$
 \\ \hline
\multicolumn{2}{c}{}
\\[0ex] 
 \hline
 States &  Non-Abelian representations of a blown-up of this tachyonic orbifold model\\
 \hline \hline 
\cellcolor{lightgray} Bosonic tachyons & \cellcolor{lightgray} none  
 \\ \hline\hline  
 Massless & $3  (\brep{10};\rep{1}) + 3  (\rep{5};\rep{1})+ 6  (\brep{5};\rep{1}) + 2  (\brep{5};\rep{1},\rep{2_+}) + 2  (\rep5;\rep2_-,\rep1) + 2 (\rep5;\rep2_+,\rep1) + (\rep5;\rep1,\rep2_-)$    
 \\
 chiral fermions & $2  (\rep1;\rep4,\rep1)+ 2  (\rep1;\rep1,\rep4)+ 2  (\rep1;\rep2_+,\rep2_+) + 4  (\rep1;\rep2_+,\rep2_-) + 2  (\rep1;\rep2_-,\rep2_+)  $
 \\ 
 & $ 4  (\rep1;\rep2_-,\rep2_-) +  6  (\rep1;\rep2_+,\rep1) + 8  (\rep1;\rep2_-,\rep1) + 34  (\rep1;\rep1,\rep2_+) + 11  (\rep1;\rep1,\rep2_-)+ 53  (\rep1;\rep1)  $ 
 \\
 \hline\hline  
 \cellcolor{lightgray} Massless & \cellcolor{lightgray}  $ (\brep{10};\rep{1}) + 9  (\rep{5};\rep{1})+ 2 (\brep{5};\rep{1}) + 2  (\rep1;\rep4,\rep1)+ 2  (\rep1;\rep1,\rep4)$
 \\ 
\cellcolor{lightgray} complex scalars & \cellcolor{lightgray}  $4  (\rep1;\rep2_+,\rep2_+) + 2  (\rep1;\rep2_+,\rep2_-) + 4  (\rep1;\rep2_-,\rep2_+) + 2  (\rep1;\rep2_-,\rep2_-) + 43 (\rep1;\rep1) $
 \\ \hline
 \end{tabular}
 }
\renewcommand{\arraystretch}{1}
\end{center}
 \caption{\label{tab:TachyonicOrbifoldBlowup}
 The upper table gives the spectrum of a tachyonic  $\Intr_\text{6-I}$ orbifold model. The low table gives the spectrum of a full blow-up of this tachyonic orbifold model; no tachyonic states are present anymore. }
\end{table}

\newpage

\section{Orbifold SM-like model searches}
\label{sc:Models}

Finally, we would like to mention some results of particle physics model searches on Calabi-Yau orbifold compactifications of the non-supersymmetric heterotic SO(16)$\times$SO(16) string. The aim of this investigation was two-fold: 
\enums{
\item 
We would like to estimate of how often tachyons actually appear when they are possible in principle. 
\item 
We would like to examine whether it is possible to obtain SM-like spectra from the non-supersymmetric heterotic string. 
}
In this talk we define SM-like to mean: f
\items{ 
\item 
the gauge group contains the SM gauge group with the standard SU(5) normalization of the non-anomalous hypercharge $Y$, 
\item 
a net number of three generations of chiral fermions, 
\item 
at least one Higgs scalar field, 
\item
and possibly vector-like exotic fermions w.r.t.\ the SM gauge group. 
}
To avoid severe over-counting we consider two orbifold models on the same orbifold geometry to be equivalent, when they have identical massless bosonic and fermionic and possibly tachyonic spectra (up to charges under Abelian factors). The results of our scans have been collected in Table~\ref{tb:OrbifoldScanOverview}. These models were obtained by implementing the SUSY breaking $\Intr_2$ orbifolding \eqref{NonSUSYE8} of the E$_8\times$E$_8$ theory in the ``Orbifolder package''~\cite{Nilles:2011aj}. 

A few comments about this table are in order: The first two columns indicate the number of inequivalent orbifold geometries a given orbifold twist admits. The third column gives the number of inequivalent models generated in our scans. The next column indicates the percentage of these models that are tachyon free. We see that even when orbifolds admit tachyons, they appear in at most 50\% of the cases. The last columns specify how many SM-like model were found and indicates whether they are one-, two- or multiple-Higgs models. Hence, this table shows there are many SM-like models that can be constructed on orbifolds; some of which might have interesting properties as recently suggested in~\cite{Hamada:2015ria}.

\begin{table}
\begin{center}
\renewcommand{\arraystretch}{1.2}
\scalebox{1}{
%
%
\begin{tabular}{|cc||r|r|r|c|c|}
\hline 
\multicolumn{2}{|c||}{Orbifold} & \multicolumn{1}{c|}{Inequivalent}  & \multicolumn{1}{c|}{Tachyon-free} &
\multicolumn{3}{c|}{SM-like tachyon-free models}
\\ 
\multicolumn{2}{|c||}{~~twist~~  \#(geom)~~} 
& \multicolumn{1}{c|}{scanned models} & \multicolumn{1}{c|}{percentage} & \multicolumn{1}{c|}{~~~~total~~~~} & one-Higgs & two-Higgs\\
\hline\hline
$\Intr_3$ & (1)           &     74,958\phantom{....} & 100\,\%\phantom{......} & 128\phantom{....} & 0 & 0 \\
\hline
$\Intr_4$ & (3)           &  1,100,336\phantom{....} & 100\,\%\phantom{......} & 12\phantom{....} & 0 & 0 \\
\hline
$\Intr_\text{6-I}$ & (2)  &    148,950\phantom{....} & 55\,\%\phantom{......} & 59\phantom{....} & 18 & 0 \\
\hline
$\Intr_\text{6-II}$ & (4) & 15,036,790\phantom{....} & 57\,\%\phantom{......} & 109\phantom{....} & 0 & 1 \\
\hline
$\Intr_\text{8-I}$ & (3)  &  2,751,085\phantom{....} & 51\,\%\phantom{......} & 24\phantom{....} & 0 & 0 \\
\hline
$\Intr_\text{8-II}$ & (2) &  4,397,555\phantom{....} & 71\,\%\phantom{......} & 187\phantom{....} & 1 & 1 \\
\hline
\hline
$\Intr_2 \times \Intr_2$ &(12) &  9,546,081\phantom{....} & 100\,\%\phantom{......} & 1,562\phantom{....} & 0 & 5 \\
\hline
$\Intr_2 \times \Intr_4$ & (10)  & 17,054,154\phantom{....} & 67\,\%\phantom{......} & 7,958\phantom{....} & 0 & 89 \\
\hline
$\Intr_3 \times \Intr_3$ & (5)  & 11,411,739\phantom{....} & 52\,\%\phantom{......} & 284\phantom{....} & 0 & 1 \\
\hline 
$\Intr_4 \times \Intr_4$ & (5)  & 15,361,570\phantom{....} & 64\,\%\phantom{......} & 2,460\phantom{....} & 0 & 6 \\
\hline
\end{tabular}
}
\renewcommand{\arraystretch}{1}
\end{center}
\caption{\label{tb:OrbifoldScanOverview} 
SM-like model searches on various $\Intr_N$ and $\Intr_M \times \Intr_N$ orbifold geometries. 
}
\end{table}

\section{Conclusions}

In this talk we have studied smooth and orbifold compactifications of the non-supersymmetric heterotic SO(16)$\times$SO(16) string. On smooth Calabi-Yau backgrounds we were able to recycle 
commonly employed techniques to determine both the fermionic and bosonic four dimensional spectra. We have argued that the non-supersymmetric theory never leads to tachyons on smooth Calabi-Yaus. However, twisted tachyons may arise on certain singular orbifolds. There is no conflict here, as one can show that these tachyons always decouple in full blow-up. In addition, 
we have performed SM-like model searches on selected orbifold geometries and found over 12,000 SM-like theories.

Let us close by mentioning some possible future directions: 
We are currently performing non-supersymmetric model searches on smooth Calabi-Yaus with line bundles~\cite{Blaszczyk:2015}. 
Given that the cosmological constant is persumably very large in non-supersymmetric models, it is important to investigate the detailed computation of the cosmological constant within string theory. Techniques to perform such computation are being revisited and further developed in~\cite{Angelantonj:2014dia,Abel:2015oxa}. This might be a first step in understanding the consequences of the destabilizing dilaton tadpole associated to the cosmological constant.  All our arguments to support the absence of tachyons are only valid to lowest order in the $\alpha'$ and $g_s$ expansions. It is therefore necessary to investigate perturbative and non-perturbative generation of tachyons beyond these simplifying approximations.

\subsection*{Acknowledgements}

We would like to thank the organizers of the conferences ``Frontiers in String Phenomenology'' Workshop, Ringberg Castle, Tegernsee, Germany, the Conference ``The string universeÕÕ, Mainz, Germany and Conference ``DISCRETE 2014'', KingÕs College London, UK for their kind invitations. 
We also would like to thank Heidelberg University for their kind hospitality. 
This work was supported by the LMUExcellent Programme.

\section*{References}
%

\bibliographystyle{iopart-num}
\bibliography{paper}

\end{document}

%% file: texdefs.tex
\DeclareMathSymbol{\mg}{\mathrel}{symbols}{"1D}

\newcommand{\gD}{\Delta}

\newcommand{\Slashed}{\hspace{-1.4ex}/\hspace{.2ex}}

\newcommand{\beq}{\begin{equation}}
\newcommand{\eeq}{\end{equation}}
\newcommand{\barr}{\begin{array}}
\newcommand{\earr}{\end{array}}
\newcommand{\equ}[1]{\begin{gather} #1 \end{gather}}

\newcommand{\items}[1]{\begin{itemize} #1 \end{itemize}}
\newcommand{\enums}[1]{\begin{enumerate} #1 \end{enumerate}}

\newcommand{\sfrac}[2]{\mbox{$\frac{#1}{#2}$}}
\newcounter{oldcounter}

\newcommand{\Intr}{\mathbbm{Z}}

\newcommand{\Real}{\mathbbm{R}}

\newcommand{\ba}[2]{\[\begin{array}{#2}\label{#1}}
\newcommand{\ea}{\end{array}\]}
\newcommand{\be}{\begin{equation}}
\newcommand{\ee}{\end{equation}}
\newcommand{\bea}{\begin{eqnarray}}
\newcommand{\eea}{\end{eqnarray}}

\newcommand{\rep}[1]{\mathbf{#1}}

\newcommand{\brep}[1]{\overline{\rep{#1}}}

\newcommand{\sm}{{\,\mbox{-}}}

%% file: ConSO16xSO16.pdf_t
\begin{picture}(0,0)%
\includegraphics{ConSO16xSO16.pdf}%
\end{picture}%
\setlength{\unitlength}{3947sp}%
\begingroup\makeatletter\ifx\SetFigFont\undefined%
\gdef\SetFigFont#1#2#3#4#5{%
  \reset@font\fontsize{#1}{#2pt}%
  \fontfamily{#3}\fontseries{#4}\fontshape{#5}%
  \selectfont}%
\fi\endgroup%
\begin{picture}(10566,2991)(568,-3619)
\put(8101,-1111){\makebox(0,0)[lb]{\smash{{\SetFigFont{20}{24.0}{\rmdefault}{\bfdefault}{\updefault}{\color[rgb]{0,0,0}supersymmetric}%
}}}}
\put(4201,-2836){\makebox(0,0)[lb]{\smash{{\SetFigFont{20}{24.0}{\rmdefault}{\bfdefault}{\updefault}{\color[rgb]{0,0,0}non-supersymmetric}%
}}}}
\put(3901,-3286){\makebox(0,0)[lb]{\smash{{\SetFigFont{20}{24.0}{\rmdefault}{\bfdefault}{\updefault}{\color[rgb]{0,0,0}SO(16)$\times$SO(16)}%
}}}}
\put(1051,-1636){\makebox(0,0)[lb]{\smash{{\SetFigFont{20}{24.0}{\rmdefault}{\bfdefault}{\updefault}{\color[rgb]{0,0,0}E$_8\times$E$_8$}%
}}}}
\put(9451,-1561){\makebox(0,0)[lb]{\smash{{\SetFigFont{20}{24.0}{\rmdefault}{\bfdefault}{\updefault}{\color[rgb]{0,0,0}strings}%
}}}}
\put(2251,-1636){\makebox(0,0)[lb]{\smash{{\SetFigFont{20}{24.0}{\rmdefault}{\bfdefault}{\updefault}{\color[rgb]{0,0,0}strings}%
}}}}
\put(6601,-3286){\makebox(0,0)[lb]{\smash{{\SetFigFont{20}{24.0}{\rmdefault}{\bfdefault}{\updefault}{\color[rgb]{0,0,0}strings}%
}}}}
\put(2326,-2611){\makebox(0,0)[lb]{\smash{{\SetFigFont{20}{24.0}{\rmdefault}{\bfdefault}{\updefault}{\color[rgb]{0,0,0}I.}%
}}}}
\put(9226,-2611){\makebox(0,0)[lb]{\smash{{\SetFigFont{20}{24.0}{\rmdefault}{\bfdefault}{\updefault}{\color[rgb]{0,0,0}II.}%
}}}}
\put(5176,-1036){\makebox(0,0)[lb]{\smash{{\SetFigFont{20}{24.0}{\rmdefault}{\bfdefault}{\updefault}{\color[rgb]{0,0,0}T-duality}%
}}}}
\put(901,-1111){\makebox(0,0)[lb]{\smash{{\SetFigFont{20}{24.0}{\rmdefault}{\bfdefault}{\updefault}{\color[rgb]{0,0,0}supersymmetric}%
}}}}
\end{picture}%